\documentstyle[12pt]{article}
  
\def\half{{\textstyle {1 \over 2}}}

\def\rchi{{\raise 2pt \hbox {$\chi$}}}
\def\rga{{\raise 2pt \hbox {$\gamma$}}}
\def\rg{{\raise 2 pt \hbox {$g$}}}

\def\susy{supersymmetry}
 
\def\<{\left\langle}
\def\>{\right\rangle}

\def\pt{\partial}
\def\eps{\epsilon}

\def\ol{\overline}

\def\lam{\lambda}

\def\ti{\tilde}

\begin{document}

\title{{\Large\bf  Wave Function of a \\ Supersymmetric FRW Model \\ with Vector 
Fields }}
\author{{\large\sf P.V. Moniz}\thanks{\sf e-mail: prlvm10@amtp.cam.ac.uk; 
PMONIZ@Delphi.com}~\thanks{{\sf URL: 
http://www.damtp.cam.ac.uk/user/prlvm10}}\\
Department of Applied Mathematics and Theoretical Physics \\
University of Cambridge \\ Silver Street, Cambridge, CB3 9EW\\ 
United Kingdom}

\date{\hspace{1cm}} 

\maketitle

\vspace{-2cm}

\begin{abstract}

A specific Friedmann-Roberstson-Walker (FRW) 
model derived from the theory of N=1 supergravity with gauged supermatter 
is considered in this paper. 
The supermatter content is restricted to a vector supermultiplet. 
The corresponding  Lorentz and supersymmetry quantum constraints are then  derived. 
Non-trivial solutions are subsquently found. A no-boundary solution is 
identified while another state may be interpreted as a wormhole solution. 

  \end{abstract}

\section{Introduction}

\indent


Research in supersymmetric quantum
 cosmology  using 
canonical methods started about 20 years or so 
\cite{1}-\cite{3}. Since then, many other papers have appeared in 
the literature   
\cite{5}-\cite{30}. Recent reviews on the subject 
of canonical quantum supergravity 
can be found in 
refs. \cite{CUP,31}.

 An important feature
of 
  N=1 supergravity is that it constitutes a ``square-root''  
  of gravity
\cite{1}. In fact, the Lorentz and supersymmetry 
constraints 
present in the canonical formulation of the theory 
induce   a set of coupled 
{\em first-order}  differential equations   
which the quantum states ought to satisfy. 
Such physical states will automatically 
 satisfy the usual Hamiltonian and momentum 
constraints as a consequence 
of 
the
 supersymmetry algebra \cite{1}-\cite{3}. 
Hence, canonical quantization 
methods may     provide us with supplementary and attractive 
insights as far as quantum supergravity theories are  concerned. 
In particular, when dealing  with (ultraviolet) 
divergences in quantum cosmology and gravity \cite{4}
and also removing Planckian masses 
induced by wormholes \cite{5,6}.

Important results 
for Bianchi class-A models were recently achieved 
within pure N=1 supergravity. 
On the one hand, the Hartle-Hawking (no-boundary) \cite{32} and  
wormhole (Hawking-Page)  \cite{33} states 
 were 
found  in the {\it same} spectrum of solutions  \cite{14,15}. This result improved previous attempts \cite{7}-\cite{13} 
where only one
 of these   states could be found, depending on the homogeneity conditions 
imposed on the gravitino \cite{9}. 
The reason was an overly restricted ansatz for 
the wave function of the universe, $\Psi$. 
More precisely, gravitational degrees of freedom have not 
been properly taken into account in the   expansion of $\Psi$  
in Lorentz invariant fermionic sectors (see ref. \cite{14,15,31} 
for more details).  
Another undesirable consequence of the ansatz used in \cite{7}-\cite{13} 
was the following: Bianchi class-A models were 
found to have  {\it no} 
physical states   but the trivial one, $\Psi =0$, 
when a cosmological constant was included \cite{16}-\cite{18}. 
 However, 
 an extension of the method 
in \cite{14,15} 
 did led in ref. \cite{19} 
to solutions of the form of  exponentials of the Chern-Simons functional.

The introduction of supermatter 
\cite{34} 
was contemplated in ref. \cite{20}--\cite{28}. 
A scalar supermultiplet, 
constituted by  complex scalar fields, $\phi,  \bar \phi$ and their 
spin$-{1 \over 2}$ partners, $\chi_A, \bar \chi_{A'}$ was considered in ref. 
\cite{5,11,21}--\cite{28}. 
A vector supermultiplet, formed by a  vector field $A^{(a)}_\mu$ and its 
supersymmetric partner,  was added   in ref. \cite{23,24}.
FRW models were  investigated  in ref. \cite{5,11,21}--\cite{26} 
 while
 a Bianchi IX model was analysed in ref. \cite{27,28}. 
Bianchi class-A models with Maxwell fields within N=2 supergravity were 
considered in ref. \cite{29,30}. 
The main results can be summarized as follows:

\begin{itemize}


\item A wormhole 
  solution was obtained in ref. \cite{5},
 but {\it not} 
 in ref. \cite{22}. We emphasize that the more general theory 
of N=1 supergravity with 
gauged supermatter \cite{34} was  employed in ref. \cite{22}. The reason for the 
discrepancy between \cite{5,22} 
 was analysed  in ref. \cite{25,26}  and related to the 
type of Lagrange multipliers and fermionic derivative ordering that 
were used. 

\item As far as a Hartle-Hawking state is concerned, some 
of the solutions  present in  \cite{5,11,21}--\cite{26}
bear some of the properties corresponding to the no-boundary proposal
\cite{32}. 
Unfortunately,   the supersymmetry 
constraints were  {\it not} 
 sufficient in determining the dependence of $\Psi$ with respect to 
the scalar field (cf. ref. \cite{25,26} for more details). 

\item The results found within the 
 more general  supermatter content used in ref. \cite{23,24}   were 
 disapointing:   the only allowed  physical state was $\Psi = 0$. 
 
\item Exponentials of  the N=2 supersymmetric Chern-Simons 
functional were the {\em only} type of solutions found in 
\cite{29,30} for the Bianchi class-A models.

\end{itemize}

It is certainly of interest to investigate further some of these issues. 
 On the one hand,  the apparent absence of 
wormhole solutions \cite{25,26} 
and the difficulty to  obtain  an adequate  Hartle-Hawking solution
when scalar supermultiplets are considered. On the other hand,
 why non-trivial physical states are not permited 
when all possible matter fields are present \cite{23,24}.
In addition, how can obtain other types of 
solutions different from the ones described in 
 ref.\cite{29,30}?

\vspace{0.3cm}

In  this paper 
 we will consider 
a closed  
FRW model within the theory of 
N=1 supergravity with supermatter restricted 
to a  vector supermultiplet. 
Out purpose is twofold.  First, to find and subsequently analyse 
possible solutions of the quantum constraints 
present in our model. 
Finally, we aim in providing a new perspective on the issues concerning 
\cite{23,24}.

In section 2 we will describe our 
field variables and then derive the corresponding  
Lorentz and supersymmetry constraints. 
{\it Non-trivial} solutions in different 
fermionic sectors are subsequently   obtained. We identify a {\it component} of 
the Hartle-Hawking (no-boundary) 
solution 
 \cite{32,35}. Another 
solution 
  could be interpreted as a quantum wormhole state 
\cite{33} (see also ref.  \cite{35a}).
We stress  that the Hartle-Hawking solution found here 
is 
part of the set of   solutions 
also present in ref. \cite{35}, where  a {\em non}-supersymmetric 
 FRW 
minisuperspace with Yang-Mills fields was instead 
considered. Since  N=1 
supergravity is a square-root of gravity,  our results are thus  
particularly interesting and  consistent with what should 
be expected (see also ref. \cite{1,10,11,CUP,31}).  
Our discussions and 
conclusions are present in section 3. Finally, 
we close this paper with an appendix where we 
describe and discuss the 
choice of configuration for 
 the field variables employed 
in this paper.

\section{Canonical Formulation and Quantization}

\subsection{Field Variables}

\indent

The action for our model 
 is obtained from the more general theory of 
N=1 supergravity with gauged supermatter 
present in ref. \cite{34} (see eq. (25.12)). We choose to  put all  scalar fields and corresponding 
supersymmetric partners equal to zero\footnote{ An important consequence of not having scalar fields and their fermionic partners is that 
the Killing potentials $D^{(a)}$  and all related quantitites 
are now absent \cite{31,34}.}, i.e., $\bar \phi^{I*} = \phi^I = 0$, 
$\chi_A^I = \bar \chi_{A'}^{I*} = 0$. 
Similar systems with  Yang-Mills fields coupled to N=1 supergravity
can also be found in ref. \cite{40}.
 
Our field variables will be constituted by a tetrad, 
$e^{AA'}_\mu$ (in two spinor component notation\footnote{I.e., 
$e^{AA'}_\mu = e^a_\mu\sigma_a^{AA'}$ where 
$a,\mu = 0,1,2,3$ are, respectively space-time and 
Lorentz indices, while $A=0,1, A'=0',1'$ are spinor 
indices and $\sigma^a_{AA'}$ are the 
Infeld-van de Warden translation symbols --  cf.,  e.g.,  ref. \cite{3}.}), 
 gravitino fields, 
$\psi^A_\mu, \bar \psi^{A'}_\nu$,  (where a bar denotes Hermitian conjugation), 
a vector field, $A^{(a)}_\mu$, (where $(a)$ is a group index) and the the 
corresponding spin-${1 \over 2}$  partners, $\lambda^{(a)}_A, \bar \lambda^{(a)}_{A'}$.

The restriction of this theory to a closed FRW model requires the 
introduction of  adequate  ans\"atze for the fields mentioned above.
These are discussed  in more detail in the Appendix.
The tetrad can be written as 
\begin{equation}
 e_{a\mu} = \left(\begin{array}{cc}
N (\tau) &0  \\
0 & a (\tau) E_{\hat a i} \end{array}\right)~,
~e^{a \mu} = \left( \begin{array}{cc}
N (\tau)^{-1} & 0 \\
0 & a (\tau)^{-1} E^{\hat a i} \end{array} \right)~,
\label{eq:1}
\end{equation}
where $ \hat a $ and $ i $ run from 1 to 3.  
$ E_{\hat a i} $ is a basis of left-invariant 1-forms on the unit $ S^3 $.

As far as the gravitino 
fields are concerned, the
Lagrange multipliers 
$
\psi^A_{~~0} $ and $ \bar\psi^{A'}_{~~0} $ are taken 
to be functions of time only. 
We  further include
\begin{equation} 
\psi^A_{~~i} = e^{AA'}_{~~~~i} \bar\psi_{A'}~, ~
\bar\psi^{A'}_{~~i} = e^{AA'}_{~~~~i} \psi_A~, \label{eq:2}
\end{equation}
where the new spinors $ \psi_A $ and $ \bar\psi_{A'} $ 
are functions of time only. 
This means we truncate the general decomposition
$ \psi^A_{~~B B'} = e_{B B'}^{~~~~i} \psi^A_{~~i} $, 
\begin{equation}
 \psi_{A B B'} = - 2 n^C_{~~B'} \rga_{A B  C} + {2 \over 3} \left( \beta_A
n_{B B'} + \beta_B n_{A B'} \right) - 
2 \varepsilon_{A B} n^C_{~~B'} \beta_C~, \label{eq:3}
\end{equation}
where $ \rga_{A B C} = \rga_{(A B C)} $,  
at spin$-\half$ mode level. I.e.,  
$\beta^A = {3 \over 4} n^{AA'} \ol\psi_{A'} \sim \ol\psi^A$.

Concerning the    matter fields, we
 will consider here  the choice employed 
in ref. \cite{35}-\cite{37} for the 
vector field $A^{(a)}_\mu$ (see also ref.
\cite{23,24}). This  configuration  is the simplest one 
that allows  vector 
fields to be present {\em consistently} 
in a closed FRW geometry. Our 
spin-1 field configuration is taken to be:
\begin{equation}
{\bf A}_{\mu}(t)~\omega^{\mu}  = 
\left(
{{f(t)}\over {4}}
\varepsilon_{(a)i(b)}{\cal T}^{(a)(b)}\right)\omega^i
~.\label{eq:45}
\end{equation}
Here $\{\omega^{\mu}\}$ denotes  
$\{\omega^{\mu}\} = \{ dt,\omega^i\}$, 
where 
$\omega^i = \hat E^i_{~\hat c} dx^{\hat c}$ 
$~(i,\hat c = 1,2,3)~$ are left-invariant  
one-forms on $S^3$, 
and ${\cal T}_{(a)(b)}$
are the generators of 
an internal group of transformations. For simplicity, we will 
restrict ourselves to the case of a 
 $SU(2)$  group, 
with $\tau_{(a)} = - \frac{1}{2} \eps_{(a)(b)(c)} {\cal T}_{(b)(c)}$ 
being the usual $SU(2)$  matrices.
The ansatz (\ref{eq:45})  
 implies $A^{(a)}_\mu$ to be paramatrized by 
a single  scalar function $f(t)$. 
FRW  cosmologies with this   ansatz are totally 
equivalent to a FRW minisuperspace with an effective 
 conformally coupled 
scalar field,  but with a quartic potential instead of a 
quadratic one. The choice (\ref{eq:45}) simplifies ~ considerably
any analysis of the Hamiltonian constraints (see \cite{35,35b,38}) 
and this constituted another  compelling argument to use it.

As 
fermionic partner for $A^{(a)}_\mu$ 
 we will use the more general choice
\begin{equation}
\lambda_A^{(a)}
= 
\lambda_A^{(a)} (t)~.
\label{eq:46}
\end{equation}
 
\subsection{ Quantum constraints and solutions}

\indent

We now proceed towards obtaining solutions of 
the 
quantum Lorentz and 
supersymmetry constraints (cf. ref. \cite{31} and references 
therein for further 
examples with other type of fields). To achieve this we need first 
to integrate the Lagrangian of N=1 supergravity with 
only vector multiplets (see ref. \cite{34,40}) 
over the angular variables of $S^3$. In this 
process we will use  the 
configurations decribed in the previous subsection for the field variables.
The next step consists 
in identifying our minisuperspace coordinates and 
canonical conjugate momenta. The presence of fermionic 
fields leads to second-class constraints (see e.g. ref \cite{wein}) and hence 
employing Dirac instead of Poisson brakets \cite{2,11}.
According to the guidelines described in \cite{24} we also 
redefine the  fermionic fields,  $ \psi_{A}(t) $ and 
$\lambda^{(a)}_{A}(t)$ 
in order to simplify the Dirac 
brackets.


 For the   $ \psi_{A}$-field we introduce, 

\begin{equation} \hat \psi_{A} = 
{\sqrt{3} \over 2^{1 \over 4}} \sigma a^{3 \over 2} \psi_{A}~, ~
\hat{\bar \psi}_{A'} = {\sqrt{3} \over 2^{1 \over 4}} \sigma a^{3 \over 2} \bar \psi_{A'} ~, 
\label{eq:54}
\end{equation}
where the conjugate momenta are 

\begin{equation} \pi_{\hat{ \psi}_{A}} = in_{AA'} \hat{\bar \psi}^{A'} ~,~
\pi_{\hat{\bar  \psi}_{A'}} = in_{AA'} \hat \psi^{A}~.  
\label{eq:58}
\end{equation}
The  Dirac brackets then become

\begin{equation} [\hat \psi_{A} , \hat{\bar \psi}_{A'}]_{D} = in_{AA'}~.  
\label{eq:59}
\end{equation}
Similarly for the $\lambda_A^{(a)}
$ field

\begin{equation} \hat\lambda^{(a)}_{~A} = 
{\sigma a^{ 3 \over 2} \over 2^{ 1 \over 4} }\lambda^{(a)}_{~A}, ~~
\hat{\bar \lambda}^{(a)}_{~A'} = 
{\sigma a^{ 3 \over 2} \over 2^{ 1 \over 4} }\bar \lambda^{(a)}_{~A'}~, 
\label{eq:60} 
\end{equation}
giving
\begin{equation}  \pi_{\hat{ \lambda}^{(a)}_{~A}} = -in_{AA'} \hat{\bar \lambda }^{(a)A'} ~,~
\pi_{\hat{\bar \lambda}^{(a)}_{~A'}} = -in_{AA'} \hat \lambda^{(a)A}~ , \label{eq:61}
\end{equation}
with
\begin{equation}  [\hat \lambda^{(a)}_{~A}, \hat{\bar \lambda}^{(a)}_{~A'}]_{D} = - i \delta^{ab} n_{AA'} ~.
\label{eq:62}
\end{equation}
Furthermore,
\begin{equation} [a , \pi_{a}]_{D} = 1~,  ~[f, \pi_{f}]_{D} = 1,  
\label{eq:63}\end{equation}
and the rest of the brackets are zero.

It is simpler to describe the theory using only (say) unprimed spinors, and, to this end, we define
\begin{equation} \bar \psi_{A} = 2 n_{A}^{~B'} \bar \psi_{B'}~, ~
 \bar \lambda^{(a)}_{~A} = 2 n_{A}^{~B'}
 \bar \lambda^{(a)}_{~B'} ~,
\label{eq:64}
\end{equation}
with which the new Dirac brackets are 
\begin{equation}  
 [\psi_{A}, \bar \psi_{B}]_{D} = i \epsilon_{AB} ~, [\lambda^{(a)}_{~A}, \bar \lambda^{(a)}_{~A'}]_{D} =
 - i \delta^{ab} \epsilon _{AB}~. \label{eq:65} 
 \end{equation}
The rest of the brackets remain unchanged. 
Quantum mechanically, one replaces the Dirac brackets by  anti-commutators if both arguments are odd 
(O) or commutators if 
otherwise (E): 

\begin{equation} [E_{1} , E_{2}] = i [E_{1} , E_{2}]_{D} ~,~ [O , E] = i [O , E]_{D} ~,~ \{O_{1} , O_{2}\} = i [O_{1} , O_{2}]_{D} ~.
\label{eq:66}
\end{equation}
Here, we take units with $\hbar = 1 $. The only non-zero (anti-)commutator relations are:
\begin{equation} \{\lambda^{(a)}_{~A}, \lambda^{(b)}_{~B} \} = \delta^{ab} \epsilon_{AB} ~, ~
\{\psi_{A} , \bar \psi_{B}\}  = - \epsilon_{AB}~,~
 [a , \pi_{a}]   = [f, \pi_{f}] = i  ~.
\label{eq:67} \end{equation}
We chose $ \left
(\bar\lam_{A'}^{(a)},\psi_{A} , a, f\right) $ to be the coordinates of the configuration 
space, and 
$ \left(\lam_{A}^{(a)},\bar \psi_{A}, \pi_a, \pi_f\right)$ to be the momentum operators in this representation.
Hence

\begin{equation}  \lambda^{a}_{~A} \rightarrow   -{  \partial   \over   \partial   \bar \lambda^{(a)A}} ~,~
  \bar \psi_{A} \rightarrow {  \partial   \over   \partial   \psi^{A}}~,~
  \pi_{a}  \rightarrow  {  \partial   \over   \partial   a} , ~  \pi_{f} \rightarrow -i {  \partial   \over   \partial   f}~.  
\label{eq:68}
\end{equation}

Following the ordering used in ref. \cite{5}, 
 we put all the fermionic derivatives in  $S_{A}$ on the right.
 In $ \bar S_{A} $, all the fermonic
derivatives are on the left. Implementing all these redefinitions, the 
 supersymmetry constraints  have the differential operator form

\begin{eqnarray} 
S_{A} & = &  
 - {1 \over 2 \sqrt{6}} a \psi_{A} {  \partial   
\over   \partial   a}   
 -   \sqrt{3 \over 2} \sigma^{2}a^{2} \psi_{A} \nonumber \\
& - & {1 \over 8 \sqrt{6}} \psi_{B} \psi^{B} {  \partial   \over   \partial   \psi^{A}} 
 - {1 \over 4 \sqrt{6}} \psi^{C} 
\bar \lambda^{(a)}_{~C} {  \partial   \over   \partial   \bar \lambda^{(a)A}}
 \nonumber \\
  & + & {1 \over 3 \sqrt{6}} \sigma^{a}_{~AB'} \sigma^{bCC'} n_{D}^{~B'} n^{B}_{~C'} \bar \lambda^{(a)D} \psi_{C} {  \partial   \over   \partial   
 \bar \lambda^{(b)B}}  \nonumber \\
&+ & {1 \over 6 \sqrt{6}} \sigma^{a}_{~AB'} \sigma^{bBA'}
 n_{D}^{~B'} n^{E}_{~A'} \bar \lambda^{(a)D} \bar \lambda^{(b)}_{~B}
{  \partial   \over   \partial   \psi^{E}}  \nonumber \\
& -& { 1 \over 2 \sqrt{6}} \psi_{A} \bar \lambda^{(a)C} {  \partial   \over 
\partial 
\bar \lambda^{(a)C}} + {3 \over 8 \sqrt{6}} \bar 
\lambda^{a}_{~A} \lambda^{(a)C} {  \partial  
 \over   \partial   \psi^{C}}  
\nonumber \\   
  & + & \sigma^{a}_{~AA'} n^{BA'}  \bar \lambda^{(a)}_{~B} \left( -{ \sqrt{2} \over 3}  {   \partial   \over   \partial   f} + 
{1 \over 8 \sqrt{2}} (1 -(f-1)^{2}) \sigma^{2}  \right)  
\label{eq:69}
\end{eqnarray} 
and 
\begin{eqnarray}
 \bar S_{A} & = &  {1 \over 2 \sqrt{6}} a {\pt \over \pt a} {\pt \over \pt \psi^{A}} 
 - \sqrt{3 \over 2} \sigma^{2}a^{2} {\pt \over  \pt \psi^{A}} 
 \nonumber \\
&- & {1 \over 8 \sqrt{6}} \varepsilon^{BC} { \pt \over \pt \psi^{B}} {\pt \over \pt \psi^{C}} \psi_{A} + {1 \over 4 \sqrt{6}} \varepsilon^{BC} {\pt \over \pt \psi^{B}} 
{\pt \over \pt \bar \lambda^{(a)C}} \bar \lambda^{(a)}_{~A}
 \nonumber \\
&+ & {1 \over 3 \sqrt{6}}  \sigma^{aB}_{~~A'} 
\sigma^{bCC'} n_{A}^{~A'} n^{D}_{~C'} {\pt \over \pt \psi^{D}}
{\pt \over \pt \bar \lambda^{(a)B}} \bar \lambda^{(b)}_{~C} \nonumber \\
&+ &  {1\over 6 \sqrt{6}}  \sigma^{aB}_{~~A'} 
\sigma_{D}^{~bB'} n_{A}^{~A'} n^{C}_{~B'} {\pt \over \pt \bar \lambda^{(a)B}}
{\pt \over \pt \bar \lambda^{(b)C}} \psi^{D} \nonumber \\
&+ &  { 1 \over 2 \sqrt{6}} {\pt \over \pt\psi^{A}} {\pt 
\over \pt \bar \lambda^{(a)B}} \bar \lambda^{(a)B} 
+{ 3 \over 8\sqrt{6}} {\pt \over \pt \bar \lambda^{(a)B}} 
{\pt \over \pt \bar \lambda^{(a)A}} \psi^{B} \nonumber \\
&+&   n_{A}^{~A'} \sigma^{aB}_{~~A'} \left( { 2 \sqrt{2} \over 3} 
{\pt \over \pt f} + { 1 \over 4 \sqrt{2}} (1 - (f-1)^{2})
\sigma^{2} \right) {\pt \over \pt \bar \lambda^{(a)B} }. 
\label{eq:70}
\end{eqnarray}

 The Lorentz 
constraint has the form:
\begin{equation} J_{AB} = \psi_{(A} \bar\psi^{B'}n_{B)B'} - 
  \lambda^{(a)}_{(A} \bar\lambda^{(a)B'}n_{B)B'} =  0~. 
 \label{eq:76a}
 \end{equation}
 The Lorentz constraint $ J_{AB} $ implies that a physical 
wave function should be a  Lorentz scalar. 
 We can easily see that the most general form of the wave function 
is 
\begin{eqnarray} 
  \Psi = A & + & B \psi^{C} \psi_{C}  
 + d_{a}  \lambda^{(a)C} \psi_{C} + c_{ab} \bar \lambda^{(a)C}  \bar \lambda^{(b)}_{~C}
+ e_{ab}  \bar \lambda^{(a)C}  \bar \lambda^{(b)}_{~C} \psi^{D} \psi_{D}  
\nonumber \\ 
 & + & 
 c_{abc}  \bar \lambda^{(a)C}  \bar \lambda^{(b)}_{~C} \bar \lambda^{(c)D} \psi_{D}   
  + c_{abcd} \bar \lambda^{(a)C}  \bar \lambda^{(b)}_{~C}
\bar \lambda^{(c)D}  \bar \lambda^{(d)}_{~D} + d_{abcd}
\bar \lambda^{(a)C}  \bar \lambda^{(b)}_{~C} \bar \lambda^{(c)D}  \bar \lambda^{(d)}_{~D}  \psi^{E} \psi_{E}   \nonumber \\ 
& + &
  \mu_{1}  \bar \lambda^{(2)C}  \bar \lambda^{(2)}_{~C} \bar \lambda^{(3)D} \bar \lambda^{(3)}_{~D} \bar \lambda^{(1)E} \psi_{E} 
\nonumber \\
&+ & \mu_{2}  \bar \lambda^{(1)C}  \bar \lambda^{(1)}_{~C} \bar \lambda^{(3)D} \bar \lambda^{(3)}_{~D} \bar \lambda^{(2)E} \psi_{E}
+ \mu_{3}  \bar \lambda^{(1)C}  \bar \lambda^{(1)}_{~C} \bar \lambda^{(2)D} \bar \lambda^{(2)}_{~D} \bar \lambda^{(3)E} \psi_{E}   \nonumber \\
& + & F 
\bar \lambda^{(1)C}  \bar \lambda^{(1)}_{~C} \bar \lambda^{(2)D} \bar \lambda^{(2)}_{~D} \bar \lambda^{(3)E} \bar \lambda^{(3)}_{~E} 
  +   G  \bar \lambda^{(1)C}  \bar \lambda^{(1)}_{~C} \bar \lambda^{(2)D} \bar \lambda^{(2)}_{~D} \bar \lambda^{(3)E} \bar \lambda^{(3)}_{~E}
\psi^{F} \psi_{F}~.
  \label{eq:71}
 \end{eqnarray}
where $A$, $B$,...,$G$  are functions of $a$,  $f$  only.  
This Ansatz contains all allowed combinations of the fermionic fields and 
 is the most general Lorentz invariant function we can write down.

The next step is to solve the supersymmetry constraints $ S_{A} \Psi = 0 $ and $ \bar S_{A'} \Psi = 0 $. 
Since the wave function (\ref{eq:71}) is of  even order in fermionic variables 
and 
stops at order $8$, the 
 expressions  $S_{A} \Psi = 0$ and $\bar S_{A} \Psi = 0$ will be of 
   odd order in fermionic variables and stop at order $7$.
Since fermionic terms as $\psi^A, \chi^A, \psi^A \chi^C\chi_C$, etc, are 
linearly independent Grassmanian quantities, the action of the 
quantum operators (\ref{eq:69}), (\ref{eq:70}) on (\ref{eq:71}) 
(see also appendix B in ref. \cite{34}) will produce 
 ten  equations from $S_A \Psi = 0$ and another 
ten equations from $\ol S_A \Psi = 0$. 
These equations are simply 
the bosonic expressions associated with each fermionic terms 
$\psi^A, \chi^A, \psi^A \chi^C\chi_C$, etc, and 
each bosonic expression therefore 
equated to zero. 

Among 
the equations derived from  $S_A \Psi = 0$  we obtain

\begin{equation}      
-{a \over {2\sqrt{6}}} {\partial A \over \partial a} - \sqrt{{3\over 2}} \sigma^2 a^2 A = 0, \label{eq:72}     
\end{equation}          
\begin{equation}      
-{\sqrt{2} \over 3} {\partial A \over \partial f} + {1 \over {8\sqrt{2}}} \left[1 - (f - 1)^2\right] \sigma^2~A =0~.  \label{eq:73}   
\end{equation}          
These equations  correspond, respectively, to the terms linear in $\psi_A, 
\bar\lambda^{(a)}_A$. Eq. (\ref{eq:72}) and (\ref{eq:73}) give the       
dependence of $A$ on $a$ and $f$, respectively. 
Solving these equations leads to       
$A = \hat A (a) \tilde A (a)$ as         
\begin{equation}      
A = \tilde A (f) e^{-3 \sigma^2 a^2},      
\label{eq:74}\end{equation}          
\begin{equation}      
A = \hat{A}(a)  e^{{3\over 16}\sigma^2 \left(-{f^3 \over 3} + f^2\right)},      
\label{eq:75} 
\end{equation}      
A similar relation exists for the
 $\bar S_A \Psi = 0$ equations, which from the       
$\psi^A\lambda^{(1)}_E\lambda^{(1)E} \lambda^{(2)}_E\lambda^{(2)E} 
\lambda^{(3)}_E\lambda^{(3)E} $ term in $\Psi$ give for $G = \hat G (a) \tilde G (f)$        
\begin{equation}      
G = \tilde G (f) e^{3 \sigma^2 a^2},     
\label{eq:76} 
\end{equation}          
\begin{equation}      
G = \hat{G}(a)  e^{{3\over 16}\sigma^2 \left({f^3 \over 3} - f^2\right)}.      
\label{eq:77}
\end{equation}        
We notice that 
in our case study, differently to the case of ref. \cite{5,11,22}-\cite{28}, we are 
indeed 
allowed to completely determine the dependence of $A$ and $G$ with respect to 
the scale factor $a$ and (the {\em effective} conformal scalar field) $f$.

The solution (\ref{eq:76}), (\ref{eq:77}) 
is included   in the 
  Hartle-Hawking 
(no-boundary) 
solutions of ref. \cite{35},   
where a $k=1$ FRW universe with Yang-Mills fields 
was employed within a {\em non-}supersymmetric 
quantum cosmologucal point  of view. 
In fact, we basically recover solution (3.8a)   in ref. 
\cite{35} if we replace 
$f \rightarrow f + 1$. 
As it can be checked, this procedure 
constitutes the rightful choice  
according to the definitions employed in \cite{36} for $A^{(a)}_\mu$. 
Solution  (\ref{eq:76}), (\ref{eq:77})
is also associated with an anti-self-dual solution of the Euclidianized equations 
of motion (cf. ref. \cite{35,35a}). 
However, it is relevant to emphasize that   {\rm not all} 
 the solutions present in 
\cite{35} can be recovered here. In particular, the Gaussian wave function 
(\ref{eq:76}), (\ref{eq:77}), peaked around $f=1$ 
(after implementing the above transformation), 
represents only one of the components of the 
wave function in ref. \cite{35}. The   wave 
function in ref. 
\cite{35}
 is peaked around the {\rm two} 
 minima of the corresponding quartic potential. 
In our model, 
the  potential terms 
 correspond  to a ``square-root'' of the potential present in \cite{35}.

Solution (\ref{eq:74}), (\ref{eq:75})  has the features of a  
(Hawking-Page)
wormhole solution for Yang-Mills fields 
\cite{33,35a}, 
which nevertheless has not yet  been found in  ordinary quantum cosmology.
However, in spite of  
(\ref{eq:74}), (\ref{eq:75})
being regular for $a \rightarrow 0$ and damped for 
$a \rightarrow \infty$, it may not be well behaved when $f \rightarrow -\infty$.

The equations obtained from the cubic and 5-order fermionic terms in 
$S_A\Psi = 0$ and $\bar S_A \Psi = 0$
 can be dealt with by multiplying them  by 
$n_{EE'}$ and   using the relation $n_{EE'} n^{EA'} = {1 \over 2} \epsilon_{E'}^{~A'}$ 
Notice  that  the $\sigma_a$ 
matrices are linear independent  and are orthogonal to the $n$ matrix.
We would see that such equations  provide the 
  $a, f$-dependence of the remaining terms in $\Psi$.
It is important to point out 
that the dependence 
of the coefficients in $\Psi$ corresponding to cubic fermionic 
terms 
on $a$ and $\phi,\ol\phi$ is 
{\it mixed }
 throughout several equations \cite{5,11}.  However, 
in the present  FRW minisuperspace with 
vector fields, the analogous dependence in $a, f$ occurs in 
{\it separate} equations. 
The  equations for cubic and 5-order fermionic terms further 
  imply that  any possible solutions
are neither the Hartle-Hawking or a wormhole state. In fact,  
we would get $d_{(a)} \sim a^5 \hat d_{(a)}(a) \ti d_{(a)} (f)$ and similar 
expressions for the other coefficients in $\Psi$,
 with a prefactor $a^n$, $n \neq 0$. This 
behaviour has also been found in \cite{5}.  Hence, from their $a$-dependence equations
these cannot be 
either a Hartle-Hawking or wormhole state. They correspond to 
other type of solutions which could be obtained from the corresponding Wheeler-DeWitt 
equation but with completely 
different boundary conditions.  

Finally, it is righteous to notice that 
the    Dirac bracket of the  supersymmetry constraints (\ref{eq:69}), (\ref{eq:70})  
induces an expression   whose bosonic sector corresponds 
to  the 
gravitational and vector field components  of the Hamiltonian constraint 
present 
 in 
 ref. \cite{35}. 
Hence, our results 
(\ref{eq:76}), (\ref{eq:77})
are consistent (as expected) 
within the context of 
N=1 supergravity being a square-root of gravity \cite{1,CUP,31}.

\section{ Discussions and Conclusions}

\indent

Summarizing our work, 
 we  considered in this paper the canonical formulation 
of  the more general theory of $ N = 1 $
 supergravity with supermatter \cite{28,34}
subject  to a $k=+1$  FRW geometry. 
Our field variables were the graviton  and 
 gravitino fields, a 
vector field $A^{(a)}_{\mu}$ 
and corresponding fermionic partners 
 (see the Appendix for further 
comments). 
We  set the scalar fields and their 
supersymmetric partners equal to zero. 

We derived in section 2 the   constraints 
for our minisuperspace  model and  solved the Lorentz and supersymmetry constraints.
We then obtained non-trivial solutions. We found expressions 
that  
 can be interpreted as corresponding  to a wormhole (Hawking-Page) 
\cite{33}  and 
(Hartle-Hawking) no-boundary \cite{32} solutions, respectively.  
The general  wave function constructed in the way mentioned 
in the previous section accomodates naturally the expectation that the early 
universe --- earlier than an inflationary stage --- might be dominated 
by radiation and associated fermionic fields. Our results constitute an 
approach towards such a supersymmetric scenario.

 The 
Hartle-Hawking solution found here is present in 
 the set of  solutions obtained  from 
a Wheeler-DeWitt equation in 
{\em non}-supersymmetric 
 quantum cosmology (cf. ref. \cite{35}). 
That is consistent 
with our expectations, since N=1 supergravity is a   square root of 
 gravity. 
Moreover, the 
   Dirac bracket of the  supersymmetry constraints (\ref{eq:69}), (\ref{eq:70})  
induces an expression  whose bosonic sector is precisely the 
gravitational and vector field components  of the Hamiltonian constraint 
present  in 
 ref. \cite{35}. 
Hence, the fact that our constraints satisfy a 
supersymmetry algebra and, in addition, that the solutions of the 
supersymmetry constraints  are consistent with our 
expectations of canonical formulation adequately 
supports the  choices 
for the field  variables configurations.

 As far as the problem of the null result in ref. \cite{23,24} is 
concerned, we hope our results may provide a new perspective on this 
issue. In the least, we know from the present paper that physical states 
in FRW models with vector fields obtained from N=1 supergravity with 
supermatter indeed exist. Physical states also exist when solely scalar 
multiplets are concerned \cite{31}. Thus, we could expect to merge both 
situations   and hopefully obtain non-trivial 
states. It should be noticed however that so far {\rm no} analytical solution 
has been found in {\rm non}-supersymmetric FRW quantum cosmologies with 
vector {\rm and } scalar fields. 
We hope to address all these issues in a future investigation. 

In conclusion, we believe  that  the results presented here 
positively add and contribute to our 
understanding of supersymmetric quantum cosmology.
In particular, we hope this paper 
will  further motivate 
other inquisitive researchers,  
who would  subsequently 
 ameliorate  current views on the subject 
with additonal perspectives. 

\vspace{0.4cm}

Supersymmetric quantum cosmology unquestionably constitutes an active 
and challenging subject for further research (see ref. \cite{31} for an  
outlook on potential projects). Interesting issues   which remain 
open and 
we are   aiming to address are the following: 
\begin{description}
\item [a)] Obtain conserved currents from $\Psi$, 
as consequence of the Dirac-like structure of the supersymmetry 
constraints \cite{newnew};
\item[b)] Test the validity of minisuperspace 
approximation in supersymmetric quantum 
cosmology;
\item [c)] Perform the  canonical quantization of 
black-holes in N=2  supergravity.
\end{description}

\vspace{1cm}
 
{\large\bf ACKNOWLEDGEMENTS}

\vspace{0.3cm}

The author is  grateful to 
A.D.Y. Cheng and S.W. Hawking  for pleasant conversations 
and for sharing their points of view and also to O. Bertolami for useful 
comments and suggestions.
Early motivation from discussions with  O. Obr\'egon 
and 
 R. Graham   
are also acknowledged, as well as relevant feedback from 
H. Luckock.  
 This work was supported by  
JNICT/PRAXIS XXI Fellowship BPD/6095/95.


\appendix 

\section{ FRW variables for pure N=1 supergravity }

\indent 

In this Appendix we will discuss  the choice made 
for the field variables configuration 
of our FRW model. 
In particular, we will analyse how the  the 
ans\"atze (\ref{eq:1}), (\ref{eq:2}), (\ref{eq:45}), (\ref{eq:46}) are 
affected a combination of local coordinate , Lorentz, gauge    and supersymmetry 
transformations.

The ansatz for the tetrad and gravitinos are  present in eq. (\ref{eq:1}), 
(\ref{eq:2}). Notice that tetrad and gravitino are not 
affected by the action of our 
internal transformation $SU(2)$ when no scalar fields and fermionic partners 
are present. 
Using the expressions (25.14) and (25.15) in ref. \cite{34} 
or (29)-(33), (39)-(42) and (34)-(35), (36)-(38) in \cite{31} 
we can obtain (see also ref. \cite{10,11}) for the tetrad 
\begin{eqnarray}
\delta e^{AA'}_{~~~~i} &= &  \left( - N^{AB} + a^{-1} \xi^{AB}
+ i  \epsilon^{(A} \psi^{B)} \right ) e_B^{~~A'}{}_i  \nonumber \\
~&+&  \left( - \bar  N^{A'B'} + a^{-1} \bar  \xi^{A'B'} + i
\bar \epsilon^{(A'} \bar
\psi^{B')} \right) e^A_{~~B'i} + 
  {i  \over 2} \left( \epsilon_C \psi^C + \bar \epsilon_{C'} \bar \psi^{C'} \right) 
e^{AA'}_{~~~~i}~,
\label{eq:A15}
\end{eqnarray}
where $\xi^\mu, N^{AB}, \epsilon^A$ are time-dependent vectors or 
spinors parametrizing local coordinates, Lorentz and supersymmetry 
transformations. 
A relation 
as 
$\delta e^{AA'}_i =  P_1\left[e^{AA'}_\mu, \psi^A_\mu\right] 
e^{AA'}_i$,
where 
$ P_1$ is an  expression 
(spatially independent and possibly complex) 
where all spatial and spinorial indices have been contracted,   holds 
provided that the relations
\begin{equation}
N^{AB} - a^{-1} \xi^{AB} - i  \epsilon^{(A} \psi^{B)} = 0~, 
\bar N^{A'B'} - a^{-1} \bar \xi^{A'B'} - i \bar
\epsilon^{(A'} \bar \psi^{B')} = 0~, \label{eq:A16}
\end{equation}
between the generators of Lorentz, coordinate and supersymmetry 
transformations are satisfied.
Hence, we will achieve $\delta e^{AA'}_{~~i} = 
C(t) \delta e^{AA'}_{~~i}$ with  $C(t) = 
{i  \over 2} \left( \epsilon_C \psi^C + \bar \epsilon_{C'} \bar \psi^{C'} \right)$
 and the 
ansatz (\ref{eq:1}) will be 
transformed into a similar configuration. Notice that 
any Grassman-algebra-valued field can be decomposed into a ``body'' or 
component along unity (which takes values in the 
field of real or complex numbers) and a ``soul'' which is nilpotent 
(see ref. \cite{nilp}). 
The combined variation above  implies that 
$\delta e^{AA'}_{~~i}$ exists entirely in the nilpotent (``soul'') part.

Let us now address how  configuration  (\ref{eq:2}) is transformed. Under 
local coordinate, Lorentz and supersymmetry transformations (when 
$\phi=\bar\phi=0$) we get 
\begin{eqnarray}
\delta \psi^A_{~~i} & = & 
a^{-1} \bar \xi^{A'B'} e^A_{~~B'i} \bar \psi_{A'}
+ {3i \over 4} \epsilon^A \psi^B \bar \psi^{B'} e_{BB'i} 
-\frac{3i}{8} \epsilon^A \lambda^{(a)C} e_{iCC'} 
\bar\lambda^{(a)C'} - \frac{3i}{4} \epsilon_C \lambda^{(a)C} 
e^A_{~C'i} \bar\lambda^{(a)C'}~
\nonumber \\
~&+ & \left[ {2} \left( {\dot a \over a N} + {i \over a} \right) - {i  \over
2 N} \left( \psi_F \psi^F_{~~0} + \bar \psi_{F'0} \bar \psi^{F'} \right) \right] n_{BA'} 
e^{AA'}_{~~~~i} \eps^B~. 
\label{eq:A18}
\end{eqnarray}
Hence,   to recover 
 a relation like   
$\delta \psi^A_i =  
P_2 \left[e^{AA'}_\mu, \psi^A_\mu\right]
e^{AA'}_i\bar \psi_{A'}$, 
we  require $\bar\xi^{A'B'} = \xi^{AB} = 0$ (see ref. \cite{11}). 
In addition, equating the second and third terms in 
(\ref{eq:A18}) to zero gives, respectively, the 
contribution of the spin-$\half$ $\psi$ and $\lambda$-fields to the Lorentz 
constraint. We further need  to consider the term 
$\epsilon_C \lambda^{(a)C} 
e^A_{~C'i} \bar\lambda^{(a)C'}$ 
as representing a field  variable  with indices $A$ and $i$, for each 
value of $(a)$. Notice that to preserve the ansatz (\ref{eq:2}) 
 (see also ref. \cite{11}) 
we had to require 
  $n_{AB'}\eps^B\sim \bar\psi_{B'}$ 
which is not quite $\bar\psi$. Here we have to deal with the 
$\lambda, \bar\lambda$--fields 
and a similar step is necessary. 
Finally, we get 
\begin{equation}
\left[ {2  } \left( {\dot a \over a N} + {i \over a} \right) - {i  \over
2 N} \left( \psi_F \psi^F_{~~0} + \bar
 \psi_{F'0} \bar \psi^{F'} \right) \right] n_{BA'} 
 \eps^B~ \sim  P_2 \left[e^{AA'}_\mu, \psi^A_\mu\right] \bar\psi_{A'}.
\label{eq:A21}
\end{equation}
This means that the variation  $\delta \psi^A_i = D(t) \psi^A_i$  
with $D(t) = \left[ {2  } \left( {\dot a \over a N} + {i \over a} \right) - {i  \over
2 N} \left( \psi_F \psi^F_{~~0} + \bar
 \psi_{F'0} \bar \psi^{F'} \right) \right]$
  will have a component along unity 
(``body'' of Grassman algebra) and another which is nilpotent 
(the ``soul'': $ \psi_F \psi^F_{~~0} + \bar
 \psi_{F'0} \bar \psi^{F'} $).

Concerning 
the choice (\ref{eq:45}) for  $A^{(a)}_\mu $ it should be noticed that 
only non-Abelian spin-1 fields can exist consistently within 
 a $k=+1$ FRW background (see ref. \cite{35}-\cite{37}), 
whose isometry group is $SO(4)$.
  More specifically, since  the physical 
observables are to be $SO(4)$-invariant, the 
fields with gauge degrees of freedom may 
transform under $SO(4)$ if these
 transformations can be compensated by a gauge 
transformation.
The idea behind the ansatz (\ref{eq:45})   is to define a homorphism of the 
isotropy group $SO(3)$ to the gauge group. 
This homomorphism defines the gauge transformation
which, for the symmetric fields,
 compensates the action of a given $SO(3)$ rotation. 
The spin-1 field   components in the basis $\left(E^i_{\hat c} dx^{\hat c}, 
\tau_{(a)}\right)$ can be
expressed as 
\begin{equation}
A_i^{(a)} = \frac{f}{2} \delta_i^{(a)}~. 
\label{eq:A48}
\end{equation}
The local coordinate and Lorentz transformations 
will correspond  to isometries and local rotations and these have been 
compensated by gauge transformations (cf. ref. \cite{35}-\cite{37} for more 
details). Under  \susy~ transformations  we get 
\begin{equation}
\delta_{(s)} A_i^{(a)} = iaE^b_i\sigma_{bAA'} \left(
\epsilon^A\bar\lambda^{A'(a)} 
-   \lambda^{A(a)}\bar\epsilon^{A'}\right)~.
\label{eq:A49}
\end{equation}
Hence,  we need to impose the following 
condition\footnote{If we had chosen $\lambda^{(a)}_A = \lambda_A$ for 
any value of $(a)$ then we would not be able to obtain a 
consistent relation similar to (\ref{eq:A50}). Namely, such that 
$\delta A^{(1)}_1 \sim A^{(1)}_1 $ and 
$\delta A^{(2)}_1 \sim A^{(2)}_1 = 0$.}
\begin{equation}
\left\{
\begin{array}{ccl}
\sigma_{bAA'} \left(\epsilon^A\bar\lambda^{A'(a)}
-   \lambda^{A(a)}\bar\epsilon^{A'}\right)  & = & E(t)~,~~\leftarrow
\left\{
\begin{array}{c}
(a)=b=1 \\
(a)=b=2 \\
(a)=b=3\\
\end{array}
\right.
\\
\sigma_{bAA'} \left(\epsilon^A\bar\lambda^{A'(a)}
-   \lambda^{A(a)}\bar\epsilon^{A'}\right)  & =  & 0~, ~~\leftarrow (a) \neq b 
\end{array}
\right. ~,
\label{eq:A50}
\end{equation}
 where $E(t)$ is spatially independent and possibly complex, 
in order to obtain $\delta A^i_{(a)} = P_3 \left[e_{AA'\mu}, 
\psi^A_\mu,\right.$  $\left.
A^{(a)}_\mu, \lambda_A^{(a)}\right] A_{(a)}^i$.
From eq. (\ref{eq:A50}) it follows that the preservation of the 
ansatz (\ref{eq:45}) will 
require $\delta A^{(a)}_i$ to include a nilpotent (``soul'') component. 
This consequence is similar to the 
one  for the 
tetrad.

As far as the $\lam-$fields are concerned, we obtain the following result 
for 
a combined local coordinate, Lorentz, supersymmetry and gauge transformation:
\begin{eqnarray}
 \delta \lam^{(a)}_A = 
& - & \half {\cal F}_{0i}^{(a)} e_A^{~A'i}n^B_{A'}\eps_B
+ \frac{i}{2} {\cal F}_{ij}^{(a)} \varepsilon_{ijk} h^{\half} n_{AA'} 
e^{BA'k} \eps_B 
\nonumber \\
 & - & \frac{i}{4} \psi_{A0}\lambda^{(a)A'} n^B_{~A'} \eps_B 
- \frac{i}{8} \psi^C_0 n_{CC'} \bar\lam^{(a)C'} \eps_A 
- \frac{i}{4} \bar\psi^{A'}_0 \lambda_A^{(a)} n^{B}_{~A'} \eps_B 
+ \frac{i}{8} \bar\psi_{C'0} n^{CC'} \lam^{(a)}_C \eps_A 
\nonumber \\
& - & \frac{i}{4} \bar\psi^{A'} n_{AE'} \bar\lam^{(a)E'}n^B_{A'} \eps_B 
- \frac{i}{16}  \bar\psi_{E'}  \bar\lam^{(a)E'} \eps_A + \kappa^{abc}\zeta^b 
\lam^{(c)}_A 
 \nonumber \\
& -& i\bar\psi_{E'} n^{BE'} \bar\lam^{(a)A'} n_{AA'} \eps_B 
+ i\bar\psi_{E'}\bar\lam^{(a)E'}\eps_A 
-  \frac{1}{2}\psi^F\lambda^{(a)}_A\eps_F - \frac{i}{\sqrt{2}} 
\psi^C\lam^{(a)}_C\eps_A
\nonumber \\
& + & \frac{i}{8} \psi_A  \lam_E^{(a)}  \eps^A 
- \frac{i}{16} \psi_C  \lam^{(a)C} \eps_A 
+  2\eps_A \lam^{(a)}_B \psi^B + \eps_C\psi^C\lambda^{(a)}_A
~,
\label{eq:A51}
\end{eqnarray}
where
\begin{equation}
{\cal F}_{0i}^{(a)}  =  \dot{f}\delta^{(a)}_i ~,~
{\cal F}_{ij}^{(a)} = \frac{1}{4}(2f - f^2) \varepsilon_{ij(a)}~.
\label{eq:A52}
\end{equation}
The last sixth  terms in eq. 
(\ref{eq:A51}) may be put in a more suitable form in order to 
obtain  
$\delta\lam^A_{(a)} = P_4 \left[e_{AA'\mu}, \psi^A_\mu, 
A^{(a)}_\mu, \lambda_A^{(a)}\right] \lam^A_{(a)}$. 
This would require that the remaining terms 
to satisfy a further 
condition equated to zero.

The above results concerning the transformations
of  the physical 
variables are   consistent 
with a 
   FRW geometry if the mentioned  
restrictions are provided. One may ask if these new relations that have to be provided 
will themselves be invariant 
under a supersymmetry transformation or even a combination 
of supersymmetry, Lorentz and other transformations.
Using these relevant transformations 
we just obtain {\em additional}  conditions for the 
previous ones to be invariant. We can proceed with this 
process within a recursive way but  neither a  contradiction  or 
a clear  indication that the relations are invariant is produced.

However, supersymmetry will be one of the features of our model, in spite 
of the new relations that 
will be produced. In fact, 
the  constraints of our FRW model satisfy a supersymmetry algebra, i.e., 
we can see 
from (\ref{eq:65}), 
(\ref{eq:69}),
(\ref{eq:70}), 
(\ref{eq:76a}) 
 that $[S_A, \bar S_B]_D \sim H + J_{AB}$, 
which is fully consistent with supersymmetry. In addition, 
the solutions obtained
by 
solving the equations $S_A \Psi =0$ and its  Hermitian conjugate 
are  in agreement 
with 
 what it should be expected with 
N=1 supergravity being a square-root of gravity. 
Namely,  our solutions 
(\ref{eq:76}), (\ref{eq:77}) are also present in the 
set of solutions found in 
ref. \cite{36} for the case of a {\em non-}supersymmetric 
quantum FRW model with Yang-Mills fields.



\begin{thebibliography}{99}

 
 

\bibitem{1}  C. Teitelboim, Phys.~Rev.~Lett.~{\bf 38}, 1106 (1977).

\bibitem{2}  M. Pilati, Nuc. Phys. B {\bf 132}, 138 (1978).


\bibitem{3}   P.D. D'Eath, Phys.~Rev.~D {\bf 29}, 2199 (1984).



\bibitem{5} L.J. Alty, P.D. D'Eath and H.F. Dowker, Phys.~Rev.~D {\bf 46}, 4402
(1992).


\bibitem{7}  P.D. D'Eath, S.W. Hawking and O. Obreg\'on, Phys.~Lett.~{\bf 300}B, 44
(1993).

\bibitem{8}   P.D. D'Eath, Phys.~Rev.~D {\bf 48}, 713 (1993).

\bibitem{9}   R. Graham and H. Luckock, Phys. Rev. D  
{\bf 49}, R4981 (1994).




\bibitem{10}   P.D. D'Eath and D.I. Hughes, Phys.~Lett.~{\bf 214}B, 498 (1988).

\bibitem{11}   P.D. D'Eath and D.I. Hughes, Nucl.~Phys.~B {\bf 378}, 381 (1992).


\bibitem{12}   M. Asano, M. Tanimoto and N. Yoshino, Phys.~Lett.~{\bf 314}B, 303 (1993).

\bibitem{13} H. Luckock and C. Oliwa, 
Phys. Rev. {\bf D51} (1995) 5883.
 
\bibitem{14}  R. Graham and A. Csord\'as, {\it 
Nontrivial fermion states in supersymmetric minisuperspace},
in: 
Proceedings of the First Mexican School in Gravitation 
and Mathematical Physics, 
Guanajuato, Mexico, December 12-16, 1994 (gr-qc/9503054 );

 


\bibitem {15} R. Graham and A. Csord\'as, Phys. Rev. Lett. {\bf 74} (1995) 4926.





\bibitem{16} P.D. D'Eath, Phys. Lett. B{\bf 320}, 20 (1994).

\bibitem{17} A.D.Y. Cheng, P.D. D'Eath and 
P.R.L.V. Moniz, Phys. Rev. D{\bf 49} (1994) 5246.


\bibitem{18} A.D.Y. Cheng, P.D. D'Eath and 
P.R.L.V. Moniz,
{\rm Gravitation and Cosmology} 
{\bf 1} (1995) 12


\bibitem{19} R. Graham and A. Csord\'as, Phys. Rev. {\bf D52} (1995) 5653


\bibitem{20}  A.D.Y. Cheng, P.D. D'Eath and 
P.R.L.V. Moniz, 
~DAMTP-Report February R94/13,  

\bibitem{21}  A.D.Y. Cheng, P.D. D'Eath and 
P.R.L.V. Moniz,
{\rm Gravitation and Cosmology} 
{\bf 1} (1995) 1

 \bibitem{22} A.D.Y. Cheng and P.R.L.V. Moniz, 
Int. J. Mod. Phys. {\bf D4} (1995) 189
 




\bibitem{23}  A.D.Y. Cheng, P.D. D'Eath and P.R.L.V. Moniz, 
{\it Quantization of a Friedmann-Robertson-Walker model} {\it in N=1 Supergravity   with 
Gauged} {\it  Supermatter},  
in: Proceedings of the 1st Mexican School in Gravitation, Guanajuato, Mexico
December 12-16 1994,  gr-qc/9503009.

  
\bibitem{24}  A.D.Y. Cheng, P.D. D'Eath and 
P.R.L.V. Moniz, 
Class. Quantum Grav. {\bf 12} (1995) 1343-1353
 


\bibitem{25} P. Moniz, {\it The Case of the Missing Wormhole State}, in:   
Proceedings of the VI Moskow International Quantum Gravity Seminar, Moskow, Russia, 
12-19 June 1995,  to be published 
by World Scientific, DAMTP report R95/19, gr-qc/9506042, 

\bibitem{26} P. Moniz,  
Gen. Rel. Grav.  {\bf  28}  (1996) 97 

\bibitem{27} P. Moniz, {\it Back to Basics? or How can supersymmetry be used 
in a simple quantum cosmological model}, 
in: Proocedings of the 1st Mexican School in Gravitation, Guanajuato, Mexico
December 12-16 1994,  DAMTP report R95/20, gr-qc/9505002

\bibitem{28} P. Moniz,  {\it Quantization of the Bianchi type-IX model 
in N=1 Supergravity in the presence of 
 Supermatter},
 International Journal of Modern Physics {\bf A11}  (1996) 1763

 
\bibitem{29}A.D.Y. Cheng and P. Moniz, {\it Quantum Bianchi Models in N=2 Supergravity with Global O(2) Internal Symmetry}
in: Proceedings of the VI Moskow International Quantum Gravity Seminar, Moskow, Russia, 
12-19 June 1995,  to be published 
by World Scientific, DAMTP report; 

\bibitem{30}  A.D.Y. Cheng and P. Moniz, {\it Canonical Quantization of 
  Bianchi Class A Models in N=2 Supergravity}, 
-- accepted for publication in Modern Phys. Lett. {\bf A11} (1996) 227.

\bibitem{CUP} P. D'Eath, ``Supersymmetric Quantum Cosmology'', 
CUP (Cambridge, 1996)


\bibitem{31} P.V. Moniz, {\it  Supersymmetric Quantum Cosmology --- 
Shaken not Stirred}, 
 (Invited Report), 
Int. J. Mod. Physics {\bf A11} 
(1996) 4321


\bibitem{4} G. Esposito, {\it Quantum Gravity, Quantum Cosmology and Lorentzian Geometries}, 
Sprin\-ger Verlag (Berlin, 1993) and references therein.


\bibitem{6} S.W. Hawking, Phys. Rev. D{\bf 37} 904 (1988).




\bibitem{32} J.B. Hartle and S.W. Hawking, Phys.~Rev.~D {\bf 28}, 2960 (1983).


\bibitem{33}   S.W. Hawking and D.N. Page, Phys.~Rev.~D {\bf 42}, 2655 (1990).


\bibitem{34}    J. Wess and J. Bagger, {\it Supersymmetry and Supergravity},
2nd.~ed. (Princeton University Press, 1992).

 

\bibitem{35}    O. Bertolami and J.M. Mour\~ao, Class. Quantum Grav. {\bf 8} (1991) 1271; 

\bibitem{35b} O. Bertolami and P.V. Moniz, Nuc. Phys. {\bf B439} (1995) 259  

\bibitem{35a} O. Bertolami, J. Mour\~ao, R. Picken and I. Volobujev, 
Int. J. Mod. Phys. {\bf A6} (1991) 4149.

\bibitem{36}    P.V. Moniz and J. Mour\~ao, Class.  Quantum Grav. {\bf 8}, (1991) 1815;
 
\bibitem{37}      J.M. Mour\~ao,  P.V. Moniz and 
P.M. S\'a, Class.  Quantum Grav. {\bf 10} (1993) 517;


G. Gibbons and  A. Steif, Phys. Lett. { \bf B320} 245 (1994); 


M.C. Bento, O. Bertolami, J.M. Mour\~ao,  P.V. Moniz and 
P.M. S\'a, Class.  Quantum Grav. {\bf 10} (1993) 285.


\bibitem{38}    O. Bertolami, Preprint Lisbon IFM-14/90, talk presented at the XIII International Colloquium on Group Theoretical Methods in Physics, Moscow, USSR June 1990, (Springer Verlag).

\bibitem{39} O. Bertolami, J.M. Mour\~ao, R.F. Picken and I.P. Volobujev, unpublished; 
S. Shabanov, talk presented at the First Iberian Meeting on Gravity, \'Evora, 
Portugal 
September 1992, edited by M.C. Bento, O. Bertolami, J.M. Mour\~ao and 
R.F. Picken (World Scientific Press, 1993); 


see also N. Manton, Ann. Phys. {\bf 167} (1986) 328; N. Manton, 
Nuc. Phys. {\bf B193} (1981) 502.



\bibitem{40} D.Z. Freedman and J. Schwarz, Phys. Rev. {\bf D15} (1977) 1007; 

S. Ferrara, F. Gliozzi, J. Scherk and P.v. Nieuwenhuizen, Nuc. Phys. {\bf 117} (1976) 333. 


\bibitem{wein} M. Henneaux and C. Teitelboim, {\it Quantization of Gauge Systems}, 
(Princeton U.P. -- 1992).


S. Weinberg, {\em Quantum Field Theory I}, CUP, 
(Cambridge, 1996)

\bibitem{nilp} See e.g., P.C. Aichelburg and R. G\"uven, Phys. Rev. Lett. {\bf 51} 
(1983) 1613;

 P. G.O. Freund, {\it Introduction to Supersymmetry}, 
(Cambridge U.P. -- 1986); 

B.S DeWitt, {\it Supermanifolds}, (Cambridge U.P.-- 1984); 

\bibitem{41}  C. Isham and J. Nelson, Phys. Rev. {\bf D10} (1974) 3226.

\bibitem{42}   T. Christodoulakis and J. Zanelli, Phys. Lett. {\bf 102A} (1984) 227; 

T. Christodoulakis and J. Zanelli, Phys. Rev. {\bf D29} (1984) 2738; 

T. Christodoulakis and C. Papadopoulos, Phys. Rev. {\bf D38} (1988) 1063

\bibitem{43}   P. D'Eath and J.J. Halliwell, Phys.~Rev.~D {\bf 35} (1987) 1100.


\bibitem{newnew} {\em Conserved currents in supersymmetric quantum cosmology?}, 
DAMTP R96/14.

 \end{thebibliography}
\end{document}